\documentclass[english,aps,nofootinbib,longbibliography,notitlepage,superscriptaddress,prd,preprintnumbers,showkeys,twocolumn,amsmath,amssymb]{revtex4-1}

\usepackage{graphicx}
\usepackage{color}
\usepackage{hyperref}
\begin{document}


\title{The splashback radius in symmetron gravity}

\author{Omar Contigiani}
\email{contigiani@lorentz.leidenuniv.nl}
\affiliation{%
Leiden Observatory, Leiden University, PO Box 9506, Leiden 2300 RA, The Netherlands}
\affiliation{%
Lorentz Institute for Theoretical Physics, Leiden University, PO Box 9506, Leiden 2300 RA, The Netherlands
}%
\author{Valeri Vardanyan}
\email{vardanyan@lorentz.leidenuniv.nl}
\affiliation{%
Leiden Observatory, Leiden University, PO Box 9506, Leiden 2300 RA, The Netherlands}
\affiliation{%
Lorentz Institute for Theoretical Physics, Leiden University, PO Box 9506, Leiden 2300 RA, The Netherlands
}%
\author{Alessandra Silvestri}
\email{silvestri@lorentz.leidenuniv.nl}
\affiliation{%
Lorentz Institute for Theoretical Physics, Leiden University, PO Box 9506, Leiden 2300 RA, The Netherlands
}%

\date{\today}

\begin{abstract}
The splashback radius $r_\mathrm{sp}$ has been identified in cosmological $N$-body simulations as an important scale associated with gravitational collapse and the phase-space distribution of recently accreted material. We employ a semi-analytical approach to study the spherical collapse of dark matter haloes in symmetron gravity and provide insights into how the phenomenology of splashback is affected. The symmetron is a scalar-tensor theory of gravity which exhibits a screening mechanism whereby higher-density regions are screened from the effects of a fifth force. In this model, we find that, as over-densities grow over cosmic time, the inner region becomes heavily screened. In particular, we identify a sector of the parameter space for which material currently sitting at $r_\mathrm{sp}$ has followed, during the collapse, the formation of this screened region. As a result, we find that for this part of the parameter space the splashback radius is maximally affected by the symmetron force and we predict changes in $r_\mathrm{sp}$ up to around $10\%$ compared to its General Relativity value. Because this margin is within the precision of present splashback experiments, we expect this feature to soon provide constraints for symmetron gravity on previously unexplored scales.
\end{abstract}

\maketitle

\section{Introduction}
\label{sec:intro}

Gravity, one of the fundamental forces of nature, plays a crucial role in inferring  our model of the cosmos as well as all the precision constraints
placed on fundamental physics through cosmology. The theory of General Relativity (GR) introduced by Einstein a century ago \cite{Einstein:1916vd},  provided a coherent theoretical framework within which to study all gravitational phenomena. While it is arguably one of the most successful theories of modern physics, having passed a host of empirical phenomena, there remain regimes of curvature and scale where GR has yet to be accurately tested. Its theoretical and phenomenological limitations are being fully explored, with an endeavour which is carried out at virtually all energy scales, ranging from the ultraviolet properties of the theory, down to energy scale of $H_0$, associated to the present-day expansion rate  of the Universe \cite{Riess:1998cb}. 

Upcoming large scale structure (LSS) surveys will provide unprecedented constraints on gravity on cosmological scales, allowing to discriminate among many theories alternative to GR. The phenomenology of theories of modified gravity (MG)  on linear cosmological scales is fairly well understood, and it is commonly characterized in terms of modifications in the relation between matter density contrast and, respectively, the lensing and Newtonian potential \cite{Zhao:2010dz, Ade:2015rim, Pogosian:2016pwr}. On the other hand,  it is well known that non-linear mechanisms in MG theories  ``screen away'' the effects of additional degrees of freedom in high density regions. This ensures that any fifth force is suppressed and MG reduces to GR in regions where it has been tested with remarkable accuracy \cite{Will:1993ns}. 

A natural regime of interest is the intermediate range, between the screened and unscreened regimes, e.g. the regions of space at the boundaries of dark matter haloes. To this extent, a feature that is gaining prominence is the so-called \emph{splashback}, which corresponds to an observable steepening of dark matter halo density profile close to the boundary \cite{Diemer:2014xya}. Locally, the position of this steepening contains interesting information about the clustering of dark matter shells and it can be understood as the dividing radius of single-stream and multistream sectors of the dark matter phase space. This feature has already been noticed in the self-similar spherical collapse framework developed and studied in \cite{Fillmore:1984, Bertschinger:1985pd}, and generalized to 3D collapse in \cite{Lithwick:2010ej}. Self-similarity, however, is fully operational in a universe without a characteristic scale, such as the Einstein-de Sitter (EdS) universe with $\Omega_m = 1$. Even though realistic applications of the same principle to $\Lambda$CDM universe are possible \cite{Shi:2016lwp}, in this paper we will focus on the collapse in EdS scenario and will leave more realistic scenarios for a future work.       

The profiles of the largest dark matter haloes in the Universe, where galaxy clusters reside, can be mapped by measuring the deformation of background sources \cite{Kaiser:1992ps, Umetsu:2011ip}. This technique, known as lensing, has been used to measure the splashback feature around clusters \cite{Umetsu:2016cun, Contigiani:2018qxn}. It should be noted however that the most stringent constraints are obtained using the distribution of subhaloes traced by the cluster galaxy members \cite{More:2016vgs, Baxter:2017csy, Chang:2017hjt, Shin:2018pic}. In this case the interpretation is nevertheless not straightforward and an accurate comparison with $N$-body $\Lambda$CDM simulations is required. 

In this paper we consider the splashback radius in MG scenarios, investigating the microscopic effects of alternative theories of gravity on the dark matter shells accreting into the halo.  Since we aim at gaining insight on the physical details, we do not resort to numerical simulations, but rather employ a semi-analytical method based on the framework of self-similar spherical collapse of \cite{Fillmore:1984}. 
We focus on the class of theories of gravity that display the symmetron screening mechanism \cite{Hinterbichler:2010es}. While we present an overview of the symmetron gravity in the main text, let us mention here that our analysis can be easily extended to other types of screening mechanisms, e.g. to Chameleon screening exhibited by $f(R)$ models \cite{Capozziello:2003tk, Carroll:2003wy}, where the essence is the density dependence in the scalar field mass, rather than the field couplings.

We have organized our presentation as follows. In Section~\ref{sec:collapse} we introduce the self-similar density profile and present the relevant equations of motion for the collapsing shells. Section~\ref{sec:symmetron} presents the basics of  symmetron gravity. Section~\ref{sec:sph_collapse_symm} discusses our numerical methods and demonstrates the effect of the symmetron force on the phase space of the dark matter halo and the shift in the splashback radius.\footnote{In the interest of reproducibility we make our numerical codes available at \url{http://github.com/contigiani/sym-splash}. } Finally, we discuss the implications of our findings and suggest potential further studies in Section~\ref{sec:conclusions}.     

\section{Density profile}
\label{sec:collapse}
In order to study the motion of accreting material onto an overdensity, we first need to specify a matter density profile. In this work, we employ the so-called self-similar approximation to the problem of spherical collapse. In this context, the idea of self-similarity was introduced for the first time by \cite{Fillmore:1984}, where it was shown that around EdS backgrounds, where the scale factor  scales as a power-law of cosmic time, $a(t) \propto t^{2/3}$, the spherical collapse equations admit self-similar and self-consistent solutions.

Material surrounding a scale-free perturbation initially coupled to the Hubble flow eventually reaches turn-around and collapses onto a central overdensity. We denote by $R(t)$ and $M(r, t)$ the position of the turn-around radius at a time $t$ and the mass contained within the radius $r$, respectively. The mass within the turn-around radius can be written as a function of scale radius as:
\begin{equation}
	M(R, t) \propto a(t)^{s},
	 \label{eq:turnaroundM}
\end{equation}
where the parameter $s$ is referred to as the \emph{accretion rate}.  In this model, $M(R, t)$ and $R(t)$ are related to each other through
\begin{equation}
	\frac{4\pi}{3}R(t)^3 \rho_b(t) = \left(\frac{4}{3\pi}\right)^2 M(R, t),
    \label{eq:consistency}
\end{equation}
where $\rho_b(t)$ is the EdS background density at time $t$. This additionally implies that the position $R$ as a function of time also depends on $s$:
\begin{equation}
	R(t) \propto a(t)^{1+s/3}.
	\label{eq:turnaroundR}
\end{equation}
Notice that $s$ and the mass of the present-day perturbation are the only free parameters of this model. In this work, we choose a fixed value of $s=1.5$ for the accretion rate, known to be representative for the low-redshift Universe in numerical simulations \cite{Correa:2015kia, Diemer:2014xya}.

During spherical collapse, Gauss's law ensures that the trajectory for each shell of material is influenced only by the mass contained within it. The equation of motion for each shell can be written as
\begin{equation}
	\frac{d^2r}{dt^2}= -\frac{GM(r, t)}{r^2},
	\label{eq:fullorbit}
\end{equation}
where the left-hand side is the Newtonian force $F_N(r)$ proportional to Newton's gravitational constant $G$.

While before turn-around the mass within a shell is manifestly constant, afterwards this is not true: as multiple shells start orbiting the halo, their trajectories start intersecting. This phenomenon is known as shell-crossing and it is the principal reason why integrating Eq. \ref{eq:fullorbit} is not straightforward. 

If we label each shell of material by its turn-around time $t_\ast$ and radius $r_\ast$, such that $R(t_\ast) = r_\ast$, the trajectory for each shell is found to be independent from these quantities when self-similarity is satisfied. This can be verified by rewriting the equation of motion for the given shell in terms of the rescaled variables 
\begin{equation}
    \xi = \frac{r}{r_\ast},~~\tau = \frac{t}{t_\ast};
\end{equation}
and by enforcing the mass profile $M(r)$ to be of the form:
\begin{equation}
 M(r, t) = M(R, t) \mathcal{M}(r/R).
\end{equation}
Notice that, from Eq.~\ref{eq:turnaroundR} it follows that the rescaling of the local turn-around radius $\Xi = \frac{R(t)}{r_\ast}$ can be also written as a function of $\tau$ alone:
\begin{equation} 
	\Xi(\tau) = \tau^{2/3+2s/9}.
\end{equation}

\begin{figure}
    \centering
    \includegraphics[width=0.5\textwidth]{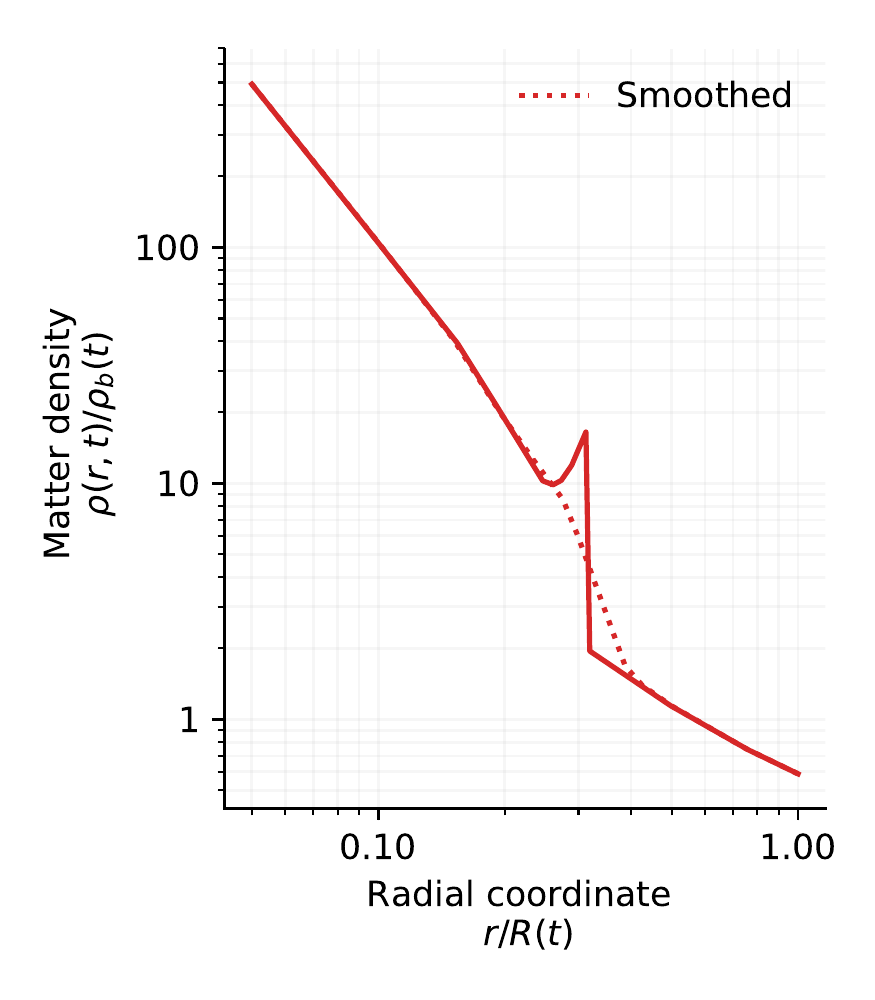}
    \caption{Prescription for the spherical halo density profile. The red dotted line is a smoothed version of the self-consistent profile which removes the non-physical sharp caustic.}
    \label{fig:rho}
\end{figure}

The system is then evolved through the following self-similarity equations for $\xi(\tau)$ and $\mathcal{M}\left( \xi/\Xi \right)$:
\begin{equation}
	\frac{d^2\xi}{d\tau^2} = - \frac{\pi^2}{8}\frac{\tau^{2s/3}}{\xi^2}
	 \mathcal{M}\left(
	 	\frac{\xi}{\Xi(\tau)}
	 \right),
	 \label{eq:xi}
\end{equation}
\begin{equation}
	\mathcal{M}(y) = \frac{2s}{3} \int_{1}^{\infty} \frac{d\tau}{\tau^{1+2s/3}} H\left( y - \frac{\xi(\tau)}{\Xi(\tau)} \right),
	\label{eq:M}
\end{equation}
where $H(\dots)$  is the Heaviside step function, and the turn-around initial conditions for $\xi(\tau)$ are $\xi(\tau = 1) = 1$, $d\xi/d\tau(\tau = 1)=0$. Notice that because these two equations are coupled to each other, they should be solved jointly to obtain self-consistent solutions for the orbits and the mass profile. This is done by starting from an initial guess for $\mathcal{M}(y)$ and then evaluating numerically the trajectories $\xi(\tau)$ using Eq.~\ref{eq:xi}. The corresponding $\mathcal{M}(y)$, evaluated using Eq.~\ref{eq:M}, is then taken as an initial guess for the next iteration. This is repeated until convergence is reached and a final result for $M(r, t)$ is obtained. The corresponding density profile is then simply

\begin{equation}
	\rho(r, t) = \frac{1}{4\pi r^2} \frac{dM}{dr}(r, t),
    \label{eq:rho_collapse}
\end{equation}

and it is shown in Figure~\ref{fig:rho}. Notice in particular that its time-dependence is completely described by $\rho_b(t)$ and $R(t)$.

\section{Symmetron gravity}
\label{sec:symmetron}
In this section we provide a brief overview of symmetron gravity and introduce the framework needed to study its effects on spherical collapse.

We consider a scalar-tensor theory of the form
	\begin{equation}
		S = S_\varphi+S_{M}\left(\tilde{g}_{\mu \nu}, \Psi \right) ,
	\end{equation}
with    
	\begin{equation}
		S_\varphi = \int \sqrt{-g} \;d^4 x  \left[  \frac{M_{p}^2}{2}R - \frac{1}{2} \nabla^\mu \varphi  \nabla_\mu \varphi - V(\varphi)\right],
	\end{equation}
$M_{p}$ being the Planck mass, and $S_\mathrm{M}$ the action for matter fields. The scalar field $\varphi$ couples to the Einstein frame metric $g_{\mu\nu}$ with Ricci scalar $R$, while matter fields (collectively represented by $\Psi$) couple to the Jordan frame metric $\tilde{g}_{\mu \nu}$. The two metrics are assumed to be related by the transformation
	\begin{equation}
		\tilde{g}_{\mu \nu} = A^2(\varphi) g_{\mu \nu}. 
	\end{equation}
	Notice that such model is fully specified by the functions $A(\varphi)$ and $V(\varphi)$. Varying the action with respect to $\varphi$ gives us the equation of motion: 
	\begin{equation}
		\square \varphi  = V_{, \varphi} - A^3 (\varphi)A_{, \varphi}(\varphi) \rho \equiv \tilde{V}_{,\varphi}(\varphi),
		\label{eq:phi}
	\end{equation}
	where $\rho$ is the trace of the matter stress-energy tensor, equal to the local matter density, and $\tilde{V}(\varphi)$ is an effective potential. The fifth force per unit mass exerted by the field $\varphi$ and experienced by a matter test particle can then be written as:
	\begin{equation}
		F_{\varphi} = -\nabla \log A(\varphi). 
		\label{eq:F}
	\end{equation}
	
	In this paper we will focus on a realization of such a theory, namely the symmetron model specified by the functions:
	\begin{equation}
		V(\varphi) = -\frac{1}{2} \mu \varphi^2 + \frac{1}{4} \lambda \varphi^4,
	\end{equation}
	\begin{equation}
		A(\varphi) = 1+ \frac{1}{2}\frac{\varphi^2}{M^2},
	\end{equation}
	and effective potential:
	\begin{equation}
		\tilde{V}(\varphi) = \frac{1}{2} \left(\frac{\rho}{M^2} - \mu^2 \right) \varphi^2 + \frac{1}{4} \lambda \varphi^4. 
		\label{eq:potential}
	\end{equation}

	In this parametrization, the symmetron naturally assumes the form of an Effective Field Theory with  ${\varphi\to -\varphi}$ symmetry. 
	
	In high density regions, where the condition	\begin{equation}
		\rho > \rho_{ssb} \equiv M^2 \mu^2
	\end{equation}
	is satisfied,  the effective potential  $\tilde{V}(\varphi)$ has only one minimum in $\varphi = 0$ and the field is driven towards it, resulting in a null fifth force. In other words,  high density regions are screened. In low density environments, on the other hand, the minimum is not located at zero. For example, for $\rho = 0$ the vacuum expectation value is $\varphi_0 = \mu / \sqrt{\lambda}$.  
	
	The fifth force can be constrained by local tests of gravity; to see in detail how local limits translate into bounds on the mass scale $M$ and the Mexican hat parameters $\mu, \lambda$ we refer the reader to~\cite{Hinterbichler:2010es}, for a general overview, and to the introduction of~\cite{OHare:2018ayv}, for a more recent analysis.
    
    In an EdS background, the average matter density as a function of redshift $z$ is
    \begin{equation}
        \rho_b = \frac{1}{6\pi G t^2} \propto (1+z)^{3}.
    \end{equation}
    As the Universe expands, the symmetron can undergo spontaneous symmetry breaking (SSB) when ${\rho_b(z_\mathrm{ssb}) = \rho_\mathrm{ssb}}$. For more details about the cosmological evolution of the symmetron field and the allowed expansion histories we refer the reader to~\cite{Hinterbichler:2011ca, Bamba:2012yf}. Let us stress however that we are not interested in the possibility of using the field $\varphi$ to drive the late-time expansion of the Universe, but we are only interested in the additional fifth force and its effects on spherical collapse. 
    
    In this paper we will work in terms of the dimensionless field $\chi = \varphi/\varphi_0$ and symmetron parameters composed by the average matter density at symmetry breaking $\rho_{ssb}$, the vacuum Compton wavelength
	\begin{equation}
		\lambda_0 = \frac{1}{\sqrt{2}\mu},
		\label{eq:lambda_0}
	\end{equation}
	and the dimensionless coupling
	\begin{equation}
		\beta = \frac{\varphi_0 M_{p}}{M^2}.
		\label{eq:beta}
	\end{equation}
	Using these parameters, the fifth force sourced by the symmetron field can be written as:
	\begin{equation}
	    F_{\varphi} = - 16\pi G  \beta^2  \lambda_0^2\rho_{ssb} \; \chi \nabla \chi. 
        \label{eq:force}
	\end{equation}

\section{Spherical collapse with the symmetron}
\label{sec:sph_collapse_symm}
Having introduced the symmetron, let us now go back to the original goal of this paper, i.e. study spherical collapse in symmetron gravity with particular focus on splashback. 

The splashback radius is commonly defined as the point where the density profile $\rho(r)$ is at its steepest. While this steepening is noteworthy because it can be detected as a departure from an equilibrium profile, this definition is clearly not suited for our study, where we assume a predefined density profile. Fortunately, the splashback radius is also known to be connected to the apocenter of recently accreted material and the location of the latest caustic visible in the density profile. Here we study the effects of the symmetron force on splashback by using this latter definition.

Our simulation is based on a system of equations that includes the spherical collapse equations, as discussed in Sec.~\ref{sec:collapse}, coupled to the equation for the field profile of the symmetron field,  discussed in Sec.~\ref{sec:symmetron}. We start by presenting our numerical method to compute both the symmetron field profile and the additional fifth force for the assumed density profile. We then proceed to  integrate the shell equation to predict the fractional change in the splashback position in the presence of the symmetron force.

\subsection{Field profile}\label{Subsec:field_profile}

Assuming  the temporal evolution of the field to be very fast compared to the other time-scales of the problem, i.e. the Hubble timescale and that of the clustering of matter,  the dimensionless field profile $\chi(r)$ sourced by a density profile $\rho(r, t)$ satisfies the following equation:

\begin{equation}
 \frac{d^2\chi}{dr^2}+ \frac{2}{r} \frac{d\chi}{dr} = \frac{1}{2\lambda_0^2} \left[ \left( \frac{\rho(r, t)}{\rho_{ssb}} -1 \right)\chi + \chi^3 \right]\,.
\label{eq:symmetron}
\end{equation}

 This quasi-static approximation is common in the literature \cite{Davis:2011pj, Clampitt:2011mx, Brax:2012nk} and has been tested in the context of $N$-body simulations \cite{Llinares:2013jua, Noller:2013wca}. In order to provide a rough, order of magnitude justification for this assumption, let us just mention that the timescale associated to the field dynamics in vacuum is given by $\sim \lambda_0/c$. It is clear that in order for the symmetron field to be relevant for the dynamics of the spherical collapse, this $\lambda_0$ should be of the same order of magnitude as the scale of the cluster itself. The latter, of course, is several orders of magnitude smaller than $c/H_0$.

The \textit{static} symmetron equation of motion~(\ref{eq:symmetron}) is a non-linear elliptical boundary value problem, for which we set the standard boundary conditions of vanishing spatial gradient of the field at $r = 0$ and $r \rightarrow \infty$. We use a one-dimensional version of the  Newton-Gauss-Seidel relaxation method for the numerical integration of the equation. This  is a standard method used for obtaining the scalar field profiles in $N$-body simulations with modifications of gravity mentioned above. 

In practice, we discretize our $1D$ static symmetron equation of motion on a regular grid of size $h$ and use a second order discretization scheme for all the derivatives.\footnote{We have tested some outputs of our integrator against the results of version where higher order discretization schemes are employed. For our particular problem we did not encounter significant differences in performance of the integrator and performed the main analysis with the version which employs the second order scheme.} The resulting equation takes the form

\begin{equation}
    \mathcal{L} [\chi_{i+1}, \chi_{i-1};\chi_{i}] = 0,
\end{equation}
where
\begin{equation}
    \mathcal{L} [\chi_{i+1}, \chi_{i-1};\chi_{i}] \equiv \mathcal{D}_\mathrm{K}[\chi_{i+1}, \chi_{i-1};\chi_{i}] -\mathcal{D}_\mathrm{P}[\chi_{i}, \rho_i]
\end{equation}
contains the discretization of the Laplace operator
\begin{equation}
    \mathcal{D}_\mathrm{K} \equiv \frac{ \chi_{i+1}+ \chi_{i-1} -2\chi_{i} }{h^2}
    +\frac{2}{ r_i }\frac{\chi_{i+1}-\chi_{i-1}}{2h}
\end{equation}
and effective potential:
\begin{equation}
    \mathcal{D}_\mathrm{P} = \frac{1}{\lambda_0^2}\left(\left(\frac{\rho_i}{\rho_{\text{ssb}}}-1\right)\chi_i+\chi_i^3 \right).
\end{equation}

The basic idea of the relaxation methods is to find a field profile from this equation which is closer to the real solution than a randomly chosen initial guess. This step is iterated over multiple (improved) guesses labelled $\chi_n(i)$ until convergence is reached.

At a given step we define an improved (new) field profile through
\begin{equation}\label{eq:discretization_2}
    \chi^{\text{new}}(i)=\chi_n(i) - 
    \left.
        \frac{\mathcal{L}(\chi(i))}{\partial\mathcal{L}(\chi(i))/\partial\chi(i)}
    \right\vert_{\chi(i) = \chi_n(i)}.
\end{equation}
Then we use a part of this \textit{new} $\chi$ as the field profile for our next relaxation iteration:
\begin{equation}
\chi_{n+1}(i) = \omega \chi^{\text{new}} + (1-\omega)\chi_{n},
\end{equation} 
where $0<\omega\leqslant 1$ is a weight parameter with, in principle, a problem-dependent optimal value.

We employ two intuitive convergence diagnostics, where at each step we terminate the iteration if a certain parameter is within a predefined threshold. The first parameter is the residual function:
\begin{equation}
    \mathcal{R}_1 \equiv \sqrt{\sum_{i}\mathcal{L}[\chi(i+1), \chi(i-1);\chi(i)]^2},
\end{equation}
and the second one is the all-mesh average of the fractional change in the field profile. 
\begin{equation}
\mathcal{R}_2 \equiv \sqrt{\sum_{i}(\chi^{\text{new}}(i)-\chi^{\text{old}}(i))^2}.
\end{equation}

To validate our integrator and convergence thresholds we compare the numerical solution to a known analytic solution. In our case, this known solution is an exact $\tanh(r)$ field profile, for which the corresponding density profile was recovered using Eq.~\ref{eq:symmetron}. 

When solving for the density profiles plotted in Fig.~\ref{fig:rho}, we numerically evaluate the equation of motion in the range $[0, 2]$ for $r/R(t)$, where the density profile for $r \geq R(t)$ is assumed to be constant. We make sure that the arbitrary choice of the upper limit has no effect on our results by testing larger values.

\subsection{Splashback}\label{subsec:splashback}

\begin{figure*}
\includegraphics[width=1\textwidth]{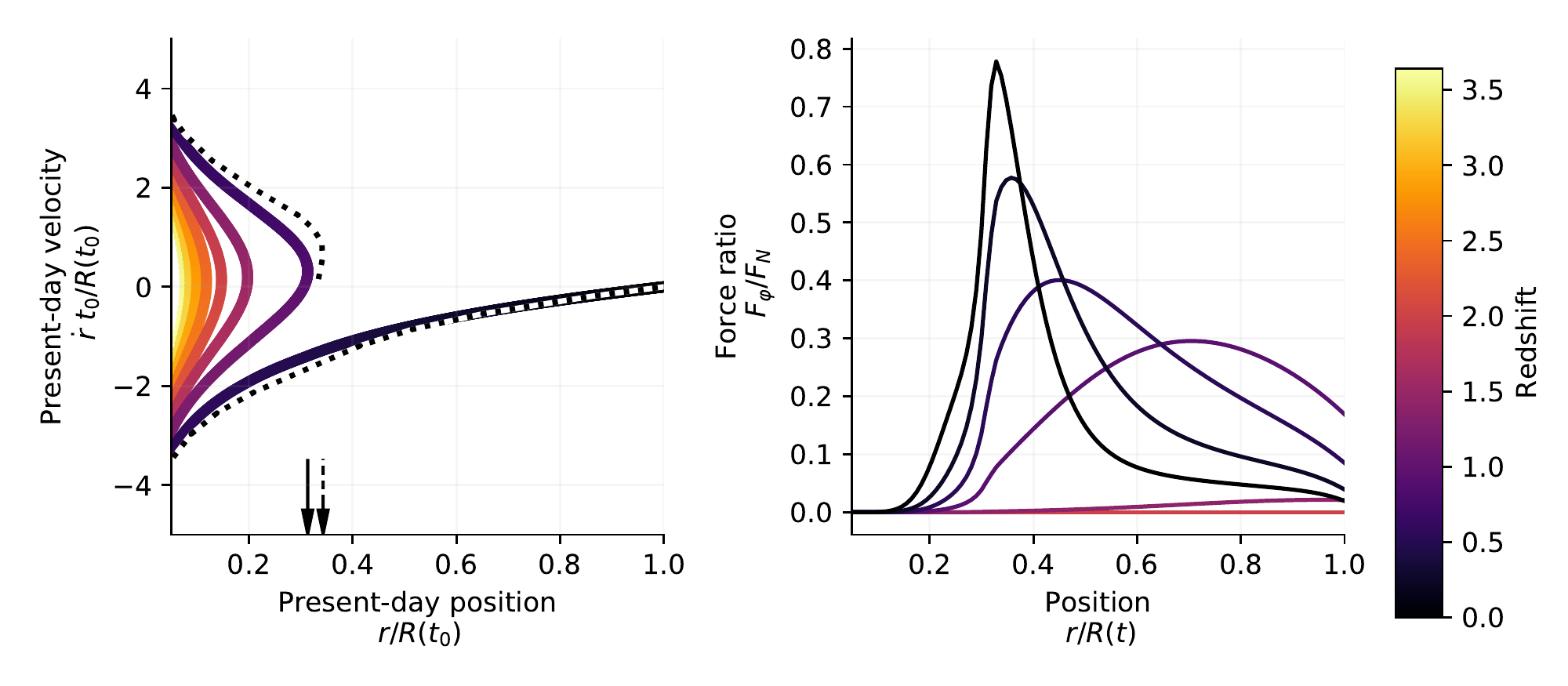}
\caption{\label{fig:example} Effects of the symmetron force on the splashback location for $\beta=3, z_\mathrm{ssb}=2, \lambda_0/R(t_0) = 0.05$. On the left side we show the phase-space distribution of shells around a spherically symmetric halo, where the shells are color-coded by their turn-around redshift. The dotted line shows how this distribution is affected by the presence of the symmetron force. The arrows on the bottom point to the inferred splashback radius in the two cases. On the right side we display the ratio between the symmetron and the Newtonian force profiles, $\frac{F_S}{F_N}$, for different instants in time. At high redshift, when the innermost material is accreted, the symmetron force is ten times smaller than its peak value today.}
\end{figure*}

Once the symmetron field is found as a function of time, the present-day phase-space distribution of recently accreted material can be obtained by integrating numerically the equation of motion~(\ref{eq:fullorbit}) with added fifth force~(\ref{eq:force}) for different collapse times.

We find that after imposing self-similarity, the collapse equations can be written only as a function of three dimensionless symmetron parameters: the redshift of symmetry breaking $z_\mathrm{ssb}$, the dimensionless coupling $\beta$, and the ratio $\lambda_0 / R(t_0)$ between the vacuum Compton wavelength $\lambda_0$  and the present-day turn-around radius $R(t_0)$. An important combination of these parameters is
\begin{equation}
    f = (1+z_\mathrm{ssb})^3 \beta^2 \frac{\lambda_0^2}{R^2(t_0)},
\end{equation}
which explicitly sets the strength of the symmetron force according to Eq. \ref{eq:F}.

From our testing, we found that values $\lambda_0/R(t_0)\in [0.02, 0.1]$ offer non-trivial cases. For $\lambda \sim R(t_0)$ we always obtain \emph{thin-shell}-like solutions, while for $\lambda \ll R(t_0)$ the field is heavy and simply relaxes onto the minimum of the potential $\tilde{V}(\chi)$ in Eq.~\ref{eq:potential}. 

In Fig.~\ref{fig:example} we illustrate our method and show how the symmetron force modifies the phase-space configuration of the latest accreted orbits (left-side plot). We find that the splashback position is  significantly affected for parameter values $f \sim 1$, $z_\mathrm{ssb}\sim2$ and $\lambda_0/R(t_0)\sim 0.1$. These values imply $M \lesssim 10^{-3} M_p$, which is in agreement with local tests of gravity \cite{Hinterbichler:2010es}.

From the same figure (right-side plot), it is clear that the innermost regions of the overdensity are screened from the effects of the fifth force at all times and this becomes relevant in the outer regions only for $z \ll z_\mathrm{ssb}$. Past this point, the force profile slowly transitions from a \emph{thick-shell} to a \emph{thin-shell} like behaviour, where the force gets progressively concentrated around the surface of the screened region \cite{Taddei:2013bsk}. Due to the sudden drop in density associated with splashback, this surface is delimited by the splashback radius. 

\begin{figure*}
    \includegraphics[width=1\textwidth]{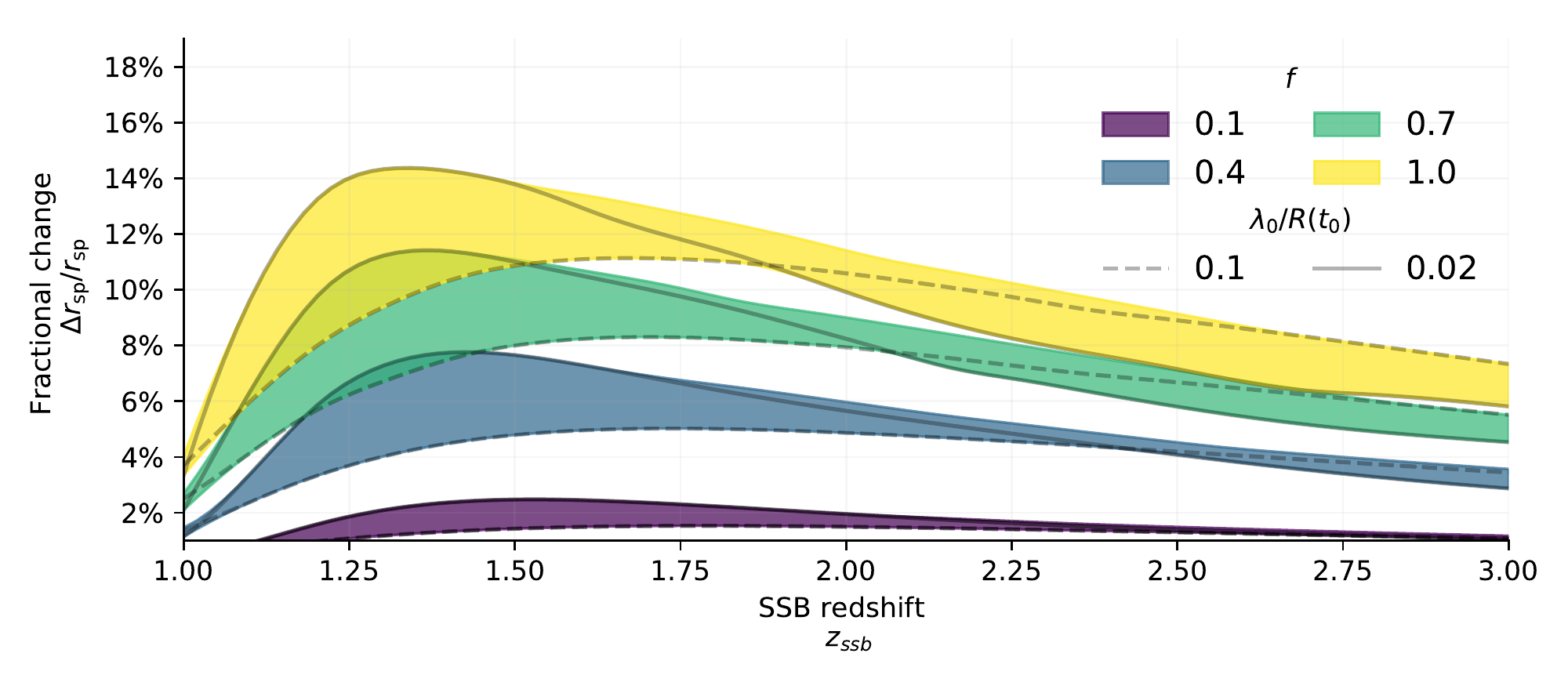}
    \caption{\label{fig:results} Percentage change in the splashback position in symmetron gravity as a function of symmetron parameters: the dimensionless force strength $f$ and the SSB redshift $z_\mathrm{ssb}$. The spread of the different curves is given by variations of the third parameter, the vacuum Compton wavelength of the field $\lambda_0$. We emphasize in particular the cases $\lambda_0/R(t_0)=0.1$ (dashed line) and $\lambda_0/R(t_0)=0.02$ (solid line), where $R(t_0)$ is the present-day turn-around radius.}
\end{figure*}

A systematic exploration of the symmetron effects on this feature as a function of all parameters is presented in Fig.~\ref{fig:results}, which represents our main result.

A clear trend with $z_\mathrm{ssb}$ is visible. Notice that the fractional change on the splashback position has an optimal peak as a function of $z_\mathrm{ssb}$ that is independent of $f$. If we call $z_\mathrm{sp}$ the accretion redshift of the shell currently sitting at the splashback position after its first pericenter, i.e. the \emph{splashback shell}, we see that the effect is maximized when $z_\mathrm{sp}\simeq z_\mathrm{ssb}$. This is easily explained by studying the profile of the fifth force over time. For $z_\mathrm{sp}\gg z_\mathrm{ssb}$, the selected shell collapses when the symmetron is in its symmetric phase and the material spends the rest of its trajectory in a screened region, away from the effects of the fifth-force; for $z_\mathrm{sp} \ll z_\mathrm{ssb}$, the thin shell has had time to form before $z_\mathrm{sp}$ and the shell feels the effects of the fifth force only during a small fraction of its trajectory. Between these two limiting cases there is an efficient $z_\mathrm{ssb}$ for which the splashback shell has time to follow the formation of the thin shell and it is optimally positioned near the peak of the force profile for most of its trajectory. In our figure we show how this peak still has a dependence on $\lambda_0$, introduced by the presence of this factor on the symmetron equation of motion~(\ref{eq:symmetron}).

To conclude this section, we point out that the smoothness of the density profile as plotted in Fig.~\ref{fig:rho} has little impact on our results and no impact on the trends discussed above. Differences between the two prescriptions exist only for $\lambda_0 \ll R(t_0)$, when the field profile becomes susceptible to the small-scale features of the profile. However, since we expect the sharp caustic to be smoothed by gravitational instabilities, for our main results we chose not to use the discontinuous profile and assumed instead its smoothed version. Notice also that considering such high-resolution scenarios would introduce additional caveats (e.g. the presence of sub-structure) that are not the focus of this work. 

\section{Discussion and conclusion}
\label{sec:conclusions}
In this work we have explored how symmetron gravity affects the splashback feature at the edges of cosmological haloes. In our approach, we assume a self-similar mass distribution motivated by spherical collapse in an EdS Universe, where the shape of the spherically symmetric matter distribution is assumed to be only a function of $r/R(t)$. This allows us to easily solve for the corresponding symmetron fifth force and estimate its effects on the splashback feature by studying the changed phase-space distribution of recently accreted shells.

The main limitation of our study is the lack of a fully consistent framework where the density profile, the turn-around physics and the phase-space distribution are solved for in conjunction with the newly introduced symmetron equation of motion. As an example, we would expect a consistent framework to take into account the back-reaction of the scalar field on the density profile. 

While deriving self-consistent solutions is outside the scope of this paper and more suited to $N$-body simulation studies, we find it useful to discuss the impact of our assumptions on the results. Changes to the turn-around physics are commonly studied through the use of different approximations, like a scale dependent Newton's constant \cite{Schaefer:2007nf, Brax:2010tj, Hu:2017aei, Nojiri:2018lwr, Lopes:2018uhq}. In our case, if we maintain the assumptions of self-similarity and power-law accretion in Eq. \ref{eq:turnaroundM}, the main change to our formalism will come in the form of upgrading the numerical constant appearing in Eq.~\ref{eq:consistency} to a function of the perturbation scale and cosmic time. 

Previous works have estimated these corrections to be of the order of a few percentage points at $z\simeq0$; see \cite{Taddei:2013bsk} for results in symmetron gravity and \cite{Lopes:2018uhq} for similar results in $f(R)$ theory. In particular, we expect our assumption to first break at a redshift $z$ such that the condition $F_\varphi(r) \sim F_N(r)$ is satisfied at the turnaround radius $r=R(t)$. In our analysis, however, we have seen that the effects on splashback are maximized when the collapse redshift of the splashback shell $z_\mathrm{sp}$ is equivalent to this transition reshift. After this point, the splashback shell is confined in the inner region and we expect its trajectory to be unaffected by the turn-around physics. Therefore, we consider our results around the peak of Fig.~\ref{fig:results} to be robust against this assumption. For the same reason, however, we expect to lose predictability for higher values of $z_\mathrm{ssb}$, since the initial condition of the splashback shell will differ from what we have assumed.

Notice that the argument presented above also implies that our results can be extended to a standard $\Lambda$CDM scenario. The present-day splashback shell is expected to have collapsed in the matter-dominated era and to have followed a trajectory mostly unaffected by the late-time expansion, especially for low values of the accretion rate $s$ like the one considered here \cite{Shi:2016lwp}.

Effects of modified gravity on the structure of dark matter haloes are usually presented in the form of changes in the small-scale power spectra \cite{Cui:2010wb, Davis:2011pj, Li:2012by, Brax:2012nk}. In this analysis we focused instead on a particular scale, the splashback radius, and showed that up to a $10\%$ change can be induced (Fig.~\ref{fig:results}). It should be pointed out that~\cite{Adhikari:2018izo} was the first work to explore how modified gravity affects the splashback position. We stress, however, that our work differs from theirs in three major aspects. First, here we focus on symmetron gravity which displays a different screening mechanism  from the chameleon or k-mouflage  explored in~\cite{Adhikari:2018izo} . Second, while their results based on $N$-body simulations represent more realistic predictions, they do not allow for a simple exploration of the theory parameter space. Third,  with our semi-analytical approach,  we are able to gain insight by obtaining quantitative results as a function of multiple theory parameters and provide an explanation for the visible trends.

Observationally, splashback can be measured predominantly around galaxy clusters,å for which the present-day turn-around radius $R(t_0)$ is of the order of a few Mpc. Our results therefore imply that this feature can be used to constrain fifth forces with vacuum Compton wavelength $\lambda_0$ just below the Mpc scale. Because measurements of splashback in the galaxy distribution around clusters have already achieved a precision below the size of our predicted effect \cite{More:2016vgs, Baxter:2017csy, Chang:2017hjt, Shin:2018pic}, we expect to soon be able to constrain not only the symmetron, but other fifth force models on similar scales.

Note in particular that, while other works have explored the possibility of constraining symmetron gravity on Mpc scales \cite{Hammami:2016npf, Gronke:2015ama}, the range considered here for $\lambda_0$ is unconstrained for this model. Thus we expect a measurement based on splashback to naturally complement other results based on laboratory experiments \cite{Burrage:2016rkv, Brax:2018zfb}, stellar and compact astrophysical objects \cite{Jain:2012tn, Brax:2013uh} or galactic disks and stellar clusters \cite{OHare:2018ayv, Llinares:2018dtu, Desmond:2018euk}. 

As the physics of splashback matures into a new cosmological observable, we expect it to play a powerful role on testing modifications of gravity, complementary to already established techniques such as those for large scale structure.

\begin{acknowledgments}
OC and VV are supported by a de Sitter Fellowship of the Netherlands Organization for Scientific Research (NWO). AS acknowledges support from the NWO and the Dutch Ministry of Education, Culture and Science (OCW), and also from the D-ITP consortium, a program of the NWO that is funded by the OCW
\end{acknowledgments}

\bibliography{symmetron}

\end{document}